\documentclass[twocolumn,prl,amssymb]{revtex4}
\def\H{\mathcal{H}}

\def\t{{\mbox{\tiny T}}}
\def\pt{{\mbox{\tiny $\Gamma$}}}
\def\c{\mathcal{C}}
\def\tr{\mbox{tr}}
\def\o{\!\otimes\!}
\newcommand{\ket}[1]{\mbox{$| #1 \rangle$}}
\newcommand{\bra}[1]{\mbox{$\langle #1 |$}}
\newcommand{\proj}[1]{\mbox{$| #1 \rangle\!\langle #1 |$}}

\newcommand{\CC}{{\mathbb{C}}}

\def\vs{\vspace{.2cm}}

\begin{document}
\title{All entangled states are useful for information processing}
\author{Llu\'{\i}s Masanes}
\affiliation{School of Mathematics, University of Bristol,
        Bristol BS8 1TW, U.K.}
\date{\today}
\begin{abstract}
The question whether all entangled states can be used as a
nonclassical resource has remained open so far. Here we provide a
conclusive answer to this problem for the case of systems shared by
two parties. We show that any entangled state $\sigma$ can enhance
the teleportation power of some other state. This holds even if the
state $\sigma$ is bound entangled.
\end{abstract}
\maketitle

\vs

It is widely said that ``entanglement is a physical resource", but
does this mean that all entangled states are resources? The
definition of entangled state is made in terms of the physical
resources needed for the {\em preparation} of the state: a state
is said to be entangled if it cannot be prepared by {\em local
(quantum) operations and classical communication} (LOCC).
Mathematically, entangled states are the ones which are not
separable \cite{sep}. This definition in terms of preparation
tells us nothing about the resources that can be extracted from an
entangled state. Until now, the existence of entangled states with
no extractable ``quantum resources'' has not been ruled out. In
this letter we prove that in the bipartite scenario such states do
not exist.

\vspace{.2cm}

One of the central ideas in quantum information theory is quantum
teleportation \cite{teleportation}. This procedure allows one to use
a bipartite quantum system in an entangled state as a quantum
channel. Remarkably, this has been experimentally implemented
\cite{tp}. If for a particular state the quality of its
corresponding teleportation channel is too low, one can sometimes do
better by allowing the procedure to fail with some probability, and
the sender/receiver know when this happens. This scenario is called
{\em conclusive teleportation}, and was introduced in \cite{Mor}.
The result of the present letter can be read as follows: any
entangled state can enhance the conclusive teleportation fidelity of
another state. But before going into this, let us recall some
results suggesting the uselessness of some entangled states.

\vspace{.2cm}

In 1989 Werner presented examples of bipartite entangled states
whose outcomes according to any local measurement can be simulated
by classical correlations \cite{Werner}, see also \cite{b}. In
1998 the Horodecki family proved the existence of {\em bound
entanglement}, that is, entangled states from which pure-state
entanglement cannot be obtained by LOCC, even with an arbitrarily
large supply of copies of the state \cite{bound-e}. They also
showed that with bound entangled states teleportation cannot be
performed better than through a classical channel, even if one
allows for conclusive teleportation \cite{mesf}.

\vspace{.2cm}

There are also results showing that some bound entangled states
are useful for several tasks \cite{HHHO,KLC,SST}, but none of them
has been proven to hold in general. In particular, in
\cite{activation}, it is presented a specific bound entangled
state capable of enhancing the fidelity of conclusive
teleportation of an other state. In the present letter this result
is generalized to all bound-entangled states.

\vspace{.2cm}

With the original teleportation protocol \cite{teleportation}, two
parties sharing a maximally-entangled state
\begin{equation}\label{me}
    \ket{\phi_d}=\frac{1}{\sqrt{d}}\, \sum_{s=1}^d \ket{s}\o\ket{s}\ ,
\end{equation}
can transmit an unknown arbitrary quantum state
$\ket{\psi}\in\CC^d$ by LOCC. If instead of $\ket{\phi_d}$ the two
parties share a noisy entangled state $\rho$, the teleportation
channel can be imperfect. One can quantify the quality of a
channel $\Gamma$ with the average fidelity of the output states
with the input states
\begin{equation}\label{fidelity}
    F_d(\Gamma)=\
    \int_{\CC^d}\! d\psi\ \tr\!\left[ \Gamma(\psi)\, \psi\right]\ ,
\end{equation}
where here and in the rest of the paper
$\psi=\ket{\psi}\!\bra{\psi}$. Note that when the shared state is
not $\phi_d$, the optimal teleportation protocol needs not be the
same as the (standard) one for $\phi_d$. The optimal conclusive
teleportation fidelity achievable with a given state $\rho$ is
considered in what follows.

\vspace{.2cm}

Given an arbitrary bipartite state $\rho$ (acting on
$\H_A\otimes\H_B$) we consider the $\CC^d \otimes \CC^d$ states
$\tilde{\rho}$ that can be obtained from $\rho$ by LOCC with some
probability. This probability can be arbitrarily small as long as
it is nonzero. This class of transformations is called
stochastic-LOCC (SLOCC). Each of these states $\tilde{\rho}$ is
the normalized output of a separable (not necessarily
trace-preserving) completely-positive map \cite{NC} with $\rho$ as
input,
\begin{equation}\label{map}
    \tilde{\rho} = \frac{\Omega( \rho)}
    {\tr\, \Omega(\rho)}\ .
\end{equation}
Separable maps (SEP) are the ones that can be written as
\begin{equation}\label{pp}
   \Omega(\rho)= \sum_k A_k\o B_k\, \rho\, A_k^\dag\o B_k^\dag \ .
\end{equation}
In our case, these matrices have the form $A_k:\H_A\rightarrow
\CC^d$ and $B_k:\H_B\rightarrow \CC^d$. The following quantity
plays a central role in our discussion. Denote by $E_d(\rho)$ the
largest overlap with the maximally-entangled state $\phi_d$ that a
state $\tilde{\rho}$ obtainable from $\rho$ by SLOCC can achieve:
\begin{equation}\label{mesf}
    E_d(\rho):=\sup_{\Omega \in \mbox{\scriptsize SEP}}
    \frac{\tr\left[\Omega(\rho)\, \phi_d\right]}
    {\tr\left[ \Omega(\rho)\right]}\ ,
\end{equation}
where the supremum is taken over all maps of the form (\ref{pp})
for which $\tr\left[ \Omega(\rho)\right]~>~0$.

\vspace{.2cm}

By definition, the quantity $E_d(\rho)$ is nonincreasing under
SLOCC processing of $\rho$, and thus, an entanglement monotone
\cite{Vidal}. One can interpret $E_d$ in the context of
single-copy entanglement distillation \cite{mesf}, as the
probability that $\rho$ ``looks'' like the maximally-entangled
state $\phi_d$ after the optimal SLOCC transformation. There is
also a relation between $E_d$ and teleportation. Denote by
$F_d(\rho)$ the optimal fidelity of conclusive teleportation that
can be achieved with a given state $\rho$. Notice that $d$ is the
dimension of the Hilbert space associated to the teleportation
channel, which has nothing to do with $\H_A\otimes\H_B$, where
$\rho$ acts. It is proven in \cite{mesf} that the optimal fidelity
of conclusive teleportation is
\begin{equation}\label{sf}
    F_d(\rho) = \frac{1+d\, E_d(\rho)}{1+d}\ .
\end{equation}
Hence, we can also interpret $E_d(\rho)$ as the teleportation
power of $\rho$.

\vspace{.2cm}

The reason for writing $E_d(\rho)$ as a supremum instead of a
maximum is because, for some states $\rho$, the set of values
$\{\tr[\tilde{\rho}\, \phi_d]$ : $\rho$ can be transformed into
$\tilde{\rho}$ by SLOCC$\}$ does not have a maximum. In such
cases, the probability of obtaining $\tilde{\rho}$ from $\rho$
goes to zero as $\tr[\tilde{\rho}\, \phi_d]$ goes to $E_d(\rho)$.
This phenomenon is called quasi-distillation in the case
$E_2(\rho)=1$, and is considered in \cite{mesf,activation}. In
\cite{distillation} it is shown that if $E_d(\rho)>1/d$ then
$\rho$ is distillable. In complement, $E_d(\rho)\geq 1/d$ holds
for any $\rho$, because the state $\ket{1}\o\ket{1}$ can be
prepared locally and its fidelity with $\phi_d$ is $1/d$.
Therefore, the range of $E_d$ is $[1/d,1]$. In what follows, the
central result of this letter is stated.

\vspace{2mm}

{\bf Theorem.} A bipartite state $\sigma$ is entangled if, and
only if, for all $d\geq 2$ and $\lambda\in[1/d,1)$, there exists a
bipartite state $\rho$ such that $E_d(\rho)\leq \lambda$ and $E_d
(\rho\otimes\sigma)
>\lambda\,$.

\vspace{.2cm}

In other words, any entangled state $\sigma$ is capable of
increasing the fidelity with $\phi_d$ of another state $\rho$.
Even if $\rho$ has initial fidelity arbitrarily close to 1. In
terms of teleportation the interpretation of the theorem is also
clear. We can chose the threshold fidelity $\lambda$ above which
we get satisfactory teleportation according to our needs (for
example, this threshold could be the teleportation fidelity
through a perfect classical channel). Now, we can consider the set
of states $\rho$ whose fidelity of conclusive teleportation is
upper bounded by the chosen threshold $\lambda$. If a state
$\sigma$ is entangled, no matter how weakly entangled it is, it
can help another state $\rho$ to achieve a fidelity strictly
larger than the threshold $\lambda$. In particular, all bipartite
bound entangled states are capable of enhancing the teleportation
power of other states. Recall that bound entanglement alone cannot
teleport at all \cite{mesf}, hence this result is quite
remarkable.

\vs{\em Proof of the theorem.} If $\sigma$ is a separable state
\cite{sep} we have that $E_d(\rho\otimes\sigma)=E_d(\rho)$ for any
$\rho$. This holds because separable states can be created by
LOCC, and $E_d$ is nonincreasing under LOCC. Let us prove the
other direction of the equivalence.

From now on $\sigma$ is an arbitrary entangled state acting on
$\H_{A_1}\otimes\H_{B_1}$. Also, $d$ and $\lambda$ are fixed to
some arbitrary values in their allowed ranges, $d\geq 2$ and
$\lambda\in[1/d,1)$. We have to show that there always exists a
state $\rho$ such that $E_d(\rho)\leq \lambda$ and
$E_d(\rho\otimes\sigma)>\lambda$. We fix $\rho$ to act on
$[\H_{A_2}\otimes\H_{A_3}]$ $\otimes$$[\H_{B_2}\otimes\H_{B_3}]$,
where $\H_{A_2}=\H_{A_1}$, $\H_{B_2}=\H_{B_1}$, and
$\H_{A_3}=\H_{B_3}=\CC^d$. From (\ref{mesf}) one can see that
$E_d(\rho)=E_d(\tau \rho)$ for any number $\tau>0$. Then, for
convenience, in the rest of the proof $\rho$ is allowed to be not
normalized. The only constraints on $\rho$ for being a state are
$\rho\succeq 0$ and $\rho\neq 0$ (we denote the operator
inequality by $\succeq$ and the real-number inequality by $\geq$).
Condition $\rho\succeq 0$ is equivalent to
\begin{equation}\label{positive}
    \tr[\rho\, P]\geq 0\quad \forall P:P\succeq 0\ ,
\end{equation}
which defines a convex cone \cite{co}.

Given a finite list of pairs of positive numbers $(x_1,y_1),\ldots
(x_n,y_n)$ the following inequality can be proven by induction:
\begin{equation}\label{ineq}
    \frac{x_1+\cdots+x_n}{y_1+\cdots+y_n}\, \leq\,
    \max_i \frac{x_i}{y_i}\ .
\end{equation}
Using it, one can see that in expression (\ref{mesf}) the supremum
is always achieved by a map $\Omega$ with only one term:
\begin{equation}\label{E}
    E_d(\rho)=\sup_{A,B}
    \frac{\tr\left[ A\o B\, \rho\, A^\dag\o B^\dag\phi_d\right]}
    {\tr\left[ A\o B\, \rho A^\dag\o B^\dag \right]}\ ,
\end{equation}
where $A,B$ are matrices of the form $A:[\H_{A_1}\otimes\CC^d]
\rightarrow \CC^d$ , $B:[\H_{B_1}\otimes\CC^d] \rightarrow \CC^d$.
With (\ref{E}) we can characterize the set of matrices $\rho$
satisfying $E_d(\rho)\leq \lambda$ by
\begin{equation}\label{set}
    \tr\!\left[ \rho\ A^\dag\o B^\dag \left(\lambda \mathbb{I} -\phi_d\right)
    A\o B \right] \geq 0\quad \forall A, B\ ,
\end{equation}
where $\mathbb{I}$ is the $d^2$-dimensional identity matrix. We are
interested in the intersection of the two cones (\ref{positive}) and
(\ref{set}):
\begin{equation}\label{}
    \c=\{\rho : \rho\succeq 0,\, E_d(\rho)\leq \lambda\}
\end{equation}
which is also a convex cone. Following \cite{co}, the dual cone of
$\c$ is
\begin{equation}\label{dual}
    \c^*=\{X : \tr[\rho\, X]\geq 0\
    \forall \rho\in\c\}\ .
\end{equation}
A version of Farkas Lemma \cite{farkas} states that each matrix
$X\in\c^*$ can be written as
\begin{equation}\label{element}
    X=\sum_k A_k^\dag\o B_k^\dag
    \left(\lambda \mathbb{I} -\phi_d\right)A_k\o B_k\ +
    \sum_s P_s\ ,
\end{equation}
where the two kinds of terms are of the form specified in
(\ref{positive}) and (\ref{set}).

Let us concentrate on the condition $E_d(\rho\otimes\sigma)
>\lambda$. Instead of computing the supremum in (\ref{E}) we
consider a particular filtering operation $\tilde{A}\otimes
\tilde{B}$, with which we obtain a lower bound on
$E_d(\rho\otimes\sigma)$. The chosen form of $\tilde{A}$ and
$\tilde{B}$ is
\begin{equation}
  \tilde{A} = \bra{\phi_{A_1 A_2}}\o\mathbb{I}_{A_3}\ ,
  \hspace{4mm}
  \tilde{B} = \bra{\phi_{B_1 B_2}}\o\mathbb{I}_{B_3}\ ,
\end{equation}
where $\ket{\phi_{A_1 A_2}}$ is the maximally entangled state
between the systems corresponding to $\H_{A_1}$ and $\H_{A_2}$
(which have the same dimension), and $\mathbb{I}_{A_3}$ is the
identity matrix acting on $\H_{A_3}$. Analogously for $\tilde{B}$.
With a little calculation one obtains the matrix equality
\begin{equation}\label{eq0}
    \tr_{\mbox{\tiny 12}}\!\left[
    \tilde{A}\o \tilde{B}\, (\rho_{\mbox{\tiny 23}} \otimes
    \sigma_{\mbox{\tiny 1}})\, \tilde{A}^\dag\o \tilde{B}^\dag
    \right] = z\, \tr_{\mbox{\tiny 2}} \left[
    \rho_{\mbox{\tiny 23}}\, \sigma_{\mbox{\tiny 2}}^\t
    \right]\ ,
\end{equation}
where $z$ is a positive number, $\sigma^\t$ stands for the
transpose of $\sigma$, and the subindexes indicate on which
Hilbert spaces every matrix acts and where the partial traces are
performed. Using (\ref{eq0}), a sufficient condition for
$E_d(\rho\otimes\sigma)>\lambda$ is
\begin{equation}\label{cond}
    \tr\!\left[\rho_{\mbox{\tiny 23}}
    \left(\sigma^\t_{\mbox{\tiny 2}} \o(\lambda\mathbb{I}
    -\phi_d)_{\mbox{\tiny 3}}\right) \right]
    <0\ .
\end{equation}
Let us show that there always exists a $\rho\in\c$ satisfying this
inequality by creating a contradiction.

Suppose that no single $\rho\in\c$ satisfies (\ref{cond}), this is
equivalent to saying that for all $\rho\in\c$
\begin{equation}\label{comp}
    \tr\!\left[\rho \left(\sigma^\t \o(\lambda\mathbb{I}-\phi_d)\right)
    \right]    \geq0\ .
\end{equation}
Then, by definition (\ref{dual}), the matrix $\sigma^\t \otimes
(\lambda\mathbb{I}-\phi_d)$ belongs to $\c^*$, and we can express
it as in (\ref{element}). One way of writing this is
\begin{equation}\label{principal}
    \sigma^\t \otimes (\lambda\mathbb{I}-\phi_d) -
    \Lambda\! \left(\lambda \mathbb{I} -\phi_d\right)
    \succeq 0\ ,
\end{equation}
where $\Lambda$ is separable and maps matrices acting on
$\CC^d\otimes\CC^d$ to matrices acting on
$\H_{A_1}\otimes\H_{B_1}\otimes\CC^d\otimes\CC^d$. The partial
trace of (\ref{principal}) over the space
$\H_{A_1}\otimes\H_{B_1}$ is like (\ref{ineq}), where $\$$ is a
separable map. Then, Lemma 1 implies that the left-hand side of
(\ref{principal}) is traceless, and a positive traceless matrix
can only by the null matrix. Therefore $\Lambda\! \left(\lambda
\mathbb{I} -\phi_d\right) = \sigma^\t \otimes
(\lambda\mathbb{I}-\phi_d)$, whose partial transposition is
\begin{equation}\label{final}
    \Lambda^\pt (\omega) =
    \sigma^{\t\pt} \otimes \omega\ ,
\end{equation}
where we have used the following notation. If $\nu$ is a bipartite
state, $\nu^\pt$ stands for its partial transposition \cite{ppt},
and $\Lambda^\pt$ is defined as $\Lambda^\pt(\nu)=
[\Lambda(\nu^\pt)]^\pt$ for all $\nu$. The map $\Lambda^\pt$ in
(\ref{final}) is completely positive and separable, hence the
matrix $\sigma^\t$ has positive partial transposition \cite{ppt}.
%Even more, it is shown in \cite{m} that $\sigma^\t$ must be separable, but this...
In Lemma 2 it is shown that a matrix $\sigma^{\t\pt}$ satisfying
(\ref{final}) must be separable. But this is in contradiction with
the initial assumption that $\sigma$ is entangled. Therefore, the
supposition that no single $\rho\in\c$ satisfies (\ref{cond}) is
false. $\Box$

\vs{\bf Lemma 1.} If $\$$ is a separable map (\ref{pp}) and
$\lambda \in [1/d,1)$ then
\begin{equation}\label{ineq}
    (\lambda\mathbb{I}-\phi_d) -
    \$\! \left(\lambda \mathbb{I} -\phi_d\right)
    \succeq 0
\end{equation}
implies
\begin{equation}\label{eq}
    \tr\!\left[ (\lambda\mathbb{I}-\phi_d) -
    \$(\lambda \mathbb{I} -\phi_d) \right]=0\ .
\end{equation}

{\em Proof.} Werner states \cite{Werner} are defined as
\begin{equation}\label{ws}
    \omega=\mu\, \omega^{-} +(1-\mu)\, \omega^+ \quad \mu\in[0,1]\ ,
\end{equation}
where $\omega^-$ and $\omega^+$ are unit-trace matrices
proportional, respectively, to the antisymmetric and symmetric
projectors acting on $\CC^d\otimes\CC^d$. One can check that
\begin{equation}\label{npt}
    \omega^\pt \succeq 0\quad \Leftrightarrow\quad \mu\leq 1/2\ .
\end{equation}
Using notation defined above we can write (\ref{ineq}) as
\begin{equation}\label{3}
    \left[\omega - \$^\pt (\omega)\right]^\pt
    \succeq 0\ ,
\end{equation}
where $\omega \propto (\lambda\mathbb{I}-\phi_d)^\pt$ is a
normalized Werner state (\ref{ws}). Due to the fact that
$\lambda<1$, the state $\omega$ in (\ref{3}) does not have a
positive partial transpose \cite{ppt}. The depolarization map is
defined as
\begin{equation}\label{depo}
    \Delta(\rho)=\int_{SU(d)} \!\!\!\!\!\!\!\!
    dU\, (U\o U)\,\rho\, (U\o U)^\dag\ ,
\end{equation}
where $dU$ represents the invariant measure on the group $SU(d)$.
The map $\Delta$ is completely-positive, separable, it always
outputs a Werner state, and leaves Werner states invariant
\cite{Werner}. The twirl map is defined as $\Delta^\pt$, and it is
also completely-positive and separable \cite{distillation}. Then,
we can apply $\Delta^\pt$ to the left-hand side of (\ref{3})
obtaining the inequality
\begin{equation}\label{4}
    \left[\omega - \tau \omega'\right]^\pt
    \succeq 0\ ,
\end{equation}
where $\tau \omega'= [\Delta \circ \$^\pt](\omega)$, and $\tau\geq
0$ is a normalization factor. By the properties of $\Delta$,
$\omega'$ is a Werner state, and can be written as $\omega'=\mu'\,
\omega^- + (1-\mu')\, \omega^+$. It is proven in \cite{SLOCC} that
the entanglement of Werner states, that is $\mu$, cannot be
increased by SLOCC, therefore
\begin{equation}\label{5}
    \mu'\leq \mu\ .
\end{equation}
Because the trace is invariant under partial transposition, from
(\ref{4}) it follows
\begin{equation}\label{6}
    1- \tau \geq 0\ .
\end{equation}
The left-hand side of (\ref{4}) is the partial transposition of an
unnormalized Werner state. Applying condition (\ref{npt}) to
(\ref{4}) we get
\begin{equation}\label{7}
    \mu-\tau\mu'\leq (1-\mu)-\tau(1-\mu')\ .
\end{equation}
The simultaneous satisfiability of (\ref{5}), (\ref{6}) and
(\ref{7}) is possible only if $\tau=1$. This implies that the
trace of the left-hand side of (\ref{4}) is zero, which is
equivalent to (\ref{eq}). $\Box$

\vs{\bf Lemma 2.} If $\omega$ is an entangled Werner state and
$\Lambda$ is a separable map such that $\Lambda(\omega)=
\eta\otimes\omega$, then $\eta$ must be a separable state.

\vs {\em Proof.} Let us separately prove the cases
$\omega=\omega^-$ and $\omega\neq\omega^-$, where $\omega^-$ is
the antisymmetric state.

Firstly consider $\omega=\omega^-$. Two parties sharing the state
$\omega^-$ can, by LOCC, obtain a singlet $\phi_2$ with some
probability. In particular, $A\o B\, \omega^- A^\dag \o B^\dag
\propto \phi_2$ when $A=B =\proj{1} +\proj{2}$. This implies that
there exists a SLOCC transformation $\Lambda'$ such that
$\Lambda'(\omega^-)= \eta\otimes\phi_2$. Consider the Schmidt
number for density matrices $K$, defined in \cite{sn}. If  $\eta$
is an entangled state then $K(\eta\otimes\phi_2) >2$. One can also
check that $K(\omega^-)=2$. But this is in contradiction with the
fact that the Schmidt number $K$ is non-increasing under SLOCC
transformations. Therefore $\eta$ must be separable.

Secondly consider $\omega\neq \omega^-$. Because the map $\Delta$
leaves Werner states invariant we can write
\begin{equation}\label{8}
    \left[(\mathbb{I}_{\mbox{\tiny 1}}\o\Delta_{\mbox{\tiny 2}})
    \circ\Lambda\circ\Delta\right](\omega) =
    \eta_{\mbox{\tiny 1}} \otimes \omega_{\mbox{\tiny 2}}\ ,
\end{equation}
where subindexes denote the system on which each map acts. It is
proven in \cite{SLOCC} that if $\$$ is a separable map and
$\omega$ is an entangled Werner state (\ref{npt}) satisfying
$[\Delta\circ\$\circ\Delta](\omega)=\omega$, then
$\Delta\circ\$\circ\Delta=\Delta$. Using this fact, one can define
the states $\eta^-$ and $\eta^+$ as
\begin{equation}\label{10}
    \left[(\mathbb{I}\o\Delta)
    \circ\Lambda\circ\Delta\right](\omega^\pm) =
    \eta^\pm \otimes \omega^\pm\ .
\end{equation}
Because $\omega^+$ and the compound map are separable, $\eta^+$
must be separable too. Apart from this $\eta^\pm$ are arbitrary.
Using representation (\ref{ws}) and linearity we have
\begin{equation}\label{11}
    \eta\otimes\omega=
    \mu\, \eta^- \otimes \omega^- +
    (1-\mu)\, \eta^+ \otimes \omega^+
    \ .
\end{equation}
Consider both sides of this equality as a vectors in the space of
hermitian matrices. Because the left-hand side is a product
vector, so must be the right-hand side. The fact that $\omega^+$
and $\omega^-$ are orthogonal implies that $\eta^-= \eta^+$, and
then $\eta=\eta^+$ is separable. $\Box$

\vspace{.2cm}

{\bf Final remarks.} The method used to prove the theorem does not
say much about the state $\rho$, whose entanglement is enhanced by
$\sigma$. But clearly, if the state $\sigma$ has a positive
partial transpose (PPT), whatever the values of $d$ and $\lambda$,
the corresponding state $\rho$ is not PPT. An other fact about
$\rho$ is that it is related to an entanglement witness \cite{KLC}
that detects $\sigma$. Therefore, the problem of finding $\rho$
given $\sigma$ is at least as hard as finding an entanglement
witness that detects $\sigma$. More concretely, the operator
\begin{equation}\label{witness}
    W=\tr_{\mbox{\tiny 2}}\left[ \rho_{\mbox{\tiny 12}}
    \, (\lambda\mathbb{I}-\phi_d)_{\mbox{\tiny 2}} \right]
\end{equation}
can be proved to be an entanglement witness by imposing
$E_d(\rho)\leq \lambda$. That $W$ detects $\sigma$ follows from
inequality (\ref{cond}). As a consequence of the theorem, the set of
witnesses of the form (\ref{witness}) is complete, in the sense that
it detects all entangled states.

\vspace{.2cm}

A state $\rho$ is said to be {\em 1-distillable} if $E_2(\rho)>1/2$.
It is known that all 1-distillable states are distillable
\cite{distillation}. It is proven in \cite{KLC} that, for each
bipartite state $\rho$ not being PPT nor 1-distillable, there exists
a PPT state $\sigma$, such that $\rho\otimes\sigma$ is
1-distillable. One can obtain a kind of dual result as a corollary
of the theorem shown above. That is, for each PPT state $\sigma$,
there exists a state not being 1-distillable $\rho$, such that
$\rho\otimes\sigma$ is 1-distillable.

\vspace{.2cm}

Concluding, the theorem proven in this paper clarifies some aspects
of entanglement theory that remained obscure before. In particular,
whether there is a way in which the entanglement present in bound
entangled states can manifest itself. With this new insight it
becomes clear that, though strong irreversible processes take place
in the preparation of bound entangled states, the pure-state
entanglement is still there, and can be used in some sense. This
shows that {\em all} quantum correlations have a distinctive
behavior, and thus, give an advantage over classical correlations.

\vs

{\bf Acknowledgments.} The author is thankful to Andrew Doherty,
Nick Jones, Barbara Kraus, Yeong Cherng Liang, Guifr\'e Vidal and
Karl Vollbrecht for discussions and suggestions on the manuscript.
This work has been supported by the U.K. EPSRC's ``IRC QIP''.


\begin{thebibliography}{99}

\bibitem{sep} A state $\rho$ acting on the Hilbert space $\H_A\otimes\H_B$
is said to be separable if it can be written as
\begin{equation}
    \rho = \sum_k\, \alpha_k\otimes\beta_k\ ,
\end{equation}
where $\alpha_k$ and $\beta_k$ are positive matrices acting
respectively on $\H_A$ and $\H_B$ \cite{Werner}.

\bibitem{Werner} R. F. Werner; Phys. Rev. A {\bf 40}, 4277 (1989).

\bibitem{teleportation} C. Bennett, G. Brassard, C. Crepeau, R.
    Jozsa, A. Peres, W. K. Wootters; Phys. Rev. Lett. {\bf 70}, 1895
    (1993).

\bibitem{tp} C. Bouwmeester, J. W. Pan, K. Mattle, M. Elbl, H.
    Weinfurter, A. Zeilinger; Nature (London) {\bf 390}, 575 (1997).

\bibitem{Mor} T. Mor; quant-ph/9608005.

\bibitem{b} J. Barrett; Phys. Rev. A, 65, 042302 (2002).

\bibitem{bound-e} M. Horodecki, P. Horodecki, R. Horodecki;
    Phys. Rev. Lett., {\bf 80}, no. 24, pp. 5239 (1998).

\bibitem{mesf} M. Horodecki, P. Horodecki, R. Horodecki;
    %``General teleportation channel, singlet fraction and quasi-distillation'',
    quant-ph/9807091.

\bibitem{HHHO} K. Horodecki, M. Horodecki, P. Horodecki, J.
    Oppenheim;
    Phys. Rev. Lett., vol. 94, no. 16, 160502 (2005).

\bibitem{KLC} B. Kraus, M. Lewenstein, J. I. Cirac; Phys. Rev. A {\bf 65}, 042327 (2002).

\bibitem{SST} P. W. Shor, J. A. Smolin, A. V. Thapliyal;
    quant-ph/0005117.

\bibitem{activation}  P. Horodecki, M. Horodecki, R. Horodecki;
    Phys. Rev. Lett. {\bf 82}, 1056 (1999).

\bibitem{NC} M. A. Nielsen, I. L. Chuang; {\em Quantum computation and
    quantum information} (Cambridge University Press, Cambridge, 2000).

\bibitem{Vidal} G. Vidal; J.Mod.Opt. {\bf 47}, 355 (2000).

\bibitem{distillation} M. Horodecki, P. Horodecki; Phys. Rev. A.
    {\bf 59}, 4206 (1999).

\bibitem{co} S. Boyd, L. Vandenberghe; {\em Convex Optimization}
(Cambridge University Press, Cambridge 2000).

\bibitem{farkas} B. D. Craven, J. J. Koliha; SIAM J. Math. Anal.
{\bf 8}, 983 (1977).

\bibitem{ppt} M. Horodecki, P. Horodecki, R. Horodecki; Phis. Lett.
A {\bf 223}, 1 (1996).

%\bibitem{m} Ll. Masanes; quantu-ph/0508071.

\bibitem{SLOCC} H. Yuang, Ll. Masanes; {\em In preparation.}

\bibitem{sn} B. M. Terhal, P. Horodecki; Phys. Rev. A {\em Rapid
Communications} {\bf 61}, 040301 (2000).

\end{thebibliography}
\end{document}